\documentclass[
 reprint,
 superscriptaddress,
 amsmath,amssymb,
 aps,
 pra,
 floatfix,
 twocolumn,
 longbibliography
]{revtex4-2}

\newcommand{\bra}[1]{\langle #1\rvert}
\newcommand{\ket}[1]{\lvert #1\rangle}

\newcommand{\op}[2]{\ket{#1} \bra{#2}}

\usepackage{graphicx}
\usepackage{dcolumn}
\usepackage{bm}
\usepackage{hyperref}

\usepackage{tikz}
\usepackage{dsfont}
\usetikzlibrary{calc}

\usepackage{comment}

\usepackage{qcircuit}

\usepackage{orcidlink}

\usepackage{amsmath}
\usepackage{upgreek}
\usepackage{physics}
\usepackage{isomath}
\usepackage{placeins}

\usepackage{listings}
\newcommand{\numuni}{r}                                 
\newcommand{\uniqub}[1]{u_{#1}}                         
\newcommand{\unimin}{\min[\uniqub{j}]}               
\newcommand{\unimax}{\max[\uniqub{j}]}               
\newcommand{\dyrange}{\frac{2\pi}{\max[1, \unimin]}}     

\newcommand{\numqub}{n}                                 
\newcommand{\qubnum}[1]{q_{#1}}                         
\newcommand{\measres}[1]{m_{#1}}
\newcommand{\highqubind}{\numqub - 1}                   
\newcommand{\measoutcome}[1]{M_{#1}}                    

\newcommand{\numthetas}{T}                              
\newcommand{\thetas}{\Theta}                                 
\newcommand{\numerators}[1]{S_{#1}}                     
\newcommand{\hightheta}{h}                              
\newcommand{\exampleden}{70}
\newcommand{\examplethetanumerators}{\{ 66, 93, 108, 123, 138 \}}

\newcommand{\den}{d}                                    
\newcommand{\adds}[1]{A_{#1}}                           
\newcommand{\gcds}[1]{G_{#1}}                           

\newcommand{\Z}{\mathcal{Z}}
\newcommand{\rotgate}[1]{e^{-i #1 \Z}}      
\newcommand{\rotgateletter}[1]{U^{#1}}                  

\newcommand{\cfi}{\mathcal{I}}

\newcommand{\titleabbrev}{RQPE}
\newcommand{\titlelong}{reductive quantum phase estimation}

\newcommand{\fractionnum}[1]{f_{#1}}

\usepackage{newfloat}
\DeclareFloatingEnvironment[name=Algorithm, placement=h]{alg}

\usepackage{enumitem}
\setenumerate[1]{label=(S\arabic*)}

\begin{document}


\title{Reductive Quantum Phase Estimation}
\author{Nicholas J.C. Papadopoulos\orcidlink{0000-0002-6357-0030}}
\thanks{Corresponding author; Nicholas.Papadopoulos@colorado.edu}
\affiliation{JILA, NIST, and Department of Physics, University of Colorado, 440 UCB, Boulder, CO 80309, USA}
\affiliation{Department of Computer Science, University of Colorado, 440 UCB, Boulder, CO 80309, USA}
\author{Jarrod  T.  Reilly\orcidlink{0000-0001-5410-089X}}
\affiliation{JILA, NIST, and Department of Physics, University of Colorado, 440 UCB, Boulder, CO 80309, USA}
\author{John Drew Wilson\orcidlink{0000-0001-6334-2460}}
\affiliation{JILA, NIST, and Department of Physics, University of Colorado, 440 UCB, Boulder, CO 80309, USA}
\author{Murray J. Holland\orcidlink{0000-0002-3778-1352}}
\affiliation{JILA, NIST, and Department of Physics, University of Colorado, 440 UCB, Boulder, CO 80309, USA}
%


\date{\today}
\begin{abstract}
Estimating a quantum phase is a necessary task in a wide range of fields of quantum science.
To accomplish this task, two well-known methods have been developed in distinct contexts, namely, Ramsey interferometry (RI) in atomic and molecular physics and quantum phase estimation (QPE) in quantum computing.
We demonstrate that these canonical examples are instances of a larger class of phase estimation protocols, which we call \titlelong\ (\titleabbrev) circuits. 
Here, we present an explicit algorithm that allows one to create an \titleabbrev\ circuit.
This circuit distinguishes an arbitrary set of phases with a fewer number of qubits and unitary applications, thereby solving a general class of quantum hypothesis testing to which RI and QPE belong.
We further demonstrate a trade-off between measurement precision and phase distinguishability, which allows one to tune the circuit to be optimal for a specific application.
\end{abstract}

{
\let\clearpage\relax
\maketitle
}

\section{Introduction} 
For over a century, physics has benefited greatly from exploiting interference effects between waves~\cite{Michelson,Jahrgang}.
In particular, the accurate estimation of a relative phase between parts of a wavefunction is a pivotal task in numerous fields of quantum physics and quantum computing.
For example, the central goal of quantum metrology is to construct experimental platforms capable of making high-precision measurements of the quantum phase that correspond to a physical parameter~\cite{Pezze,Degan,Reilly}.
Progress in this area has led to the development of quantum sensors that have then been used in a wide range of groundbreaking technologies, from atomic clocks~\cite{Chou,Bothwell} to medical devices~\cite{Aslam,Taylor}. 
In quantum information science, there exist important algorithms that seek to  calculate the quantum phase as precisely as possible in a single measurement. 
This can be used to find the eigenvalues of a unitary operator, thereby allowing one to perform computations such as matrix inversion and modular multiplication with a quantum advantage~\cite{nielsen_chuang_2010,Quintino,Cho}. 
For example, it is a crucial step in the Harrow-Hassidim-Lloyd linear system of equations algorithm~\cite{Harrow,Cai} as well as the crux of Shor's algorithm for prime factorization~\cite{nielsen_chuang_2010,Shor_1997}.

Due to the far-reaching impact of estimating a quantum phase, various techniques have been developed to accomplish this task.
For example, in quantum metrology the standard method is Ramsey interferometry (RI)~\cite{Ramsey}.
In RI, with a direct correspondence to optical interferometers, a single qubit is split into a coherent superposition, undergoes unitary encoding of a phase, and is then recombined (see Fig.~\ref{fig:current_solutions}(a)).
The alternate approach used in quantum computation is founded on Kitaev's quantum phase estimation (QPE) algorithm~\cite{Kitaev,nielsen_chuang_2010,Pezze2021QuantumPE}, which aims to determine the correct phase from a discrete set of possibilities in a single run of a multi-qubit quantum circuit.
The QPE algorithm consists of conditional rotations to estimate the quantum phase over small intervals of the Bloch sphere's equator, and the intervals become exponentially smaller as the number of qubits increases.

Many quantum algorithms have a phase estimation subroutine that ideally estimates an encoded phase $\theta$ after a single run of the circuit. 
This can be accomplished for phases where every qubit in the circuit has unit probability of being in one of the computational basis states (i.e., the bare eigenstates).
The canonical QPE algorithm consisting of $\numqub$ qubits can discriminate between a set of $2^{\numqub}$ phases evenly distributed throughout the interval $[ 0, 2 \pi )$.
However, there are important problems in quantum hypothesis testing~\cite{Audenaert2007,Audenaert2008,Calsamiglia2008,Spedalieri2014,Rossi2021}, a central pillar of modern quantum information science research, where one wants to discriminate between a certain discrete set of phases starting with a flat prior probability distribution. In these situations, the QPE circuit is excessive with potentially many unneeded qubits performing unnecessary rotations, increasing the chance of errors to occur during the algorithm through both quantum and classical noise.
Furthermore, there are simple situations in which QPE would actually require an infinite number of qubits to distinguish between two phases with certainty after a single run of the circuit, such is the case with $\theta = 0$ and $\theta = \pi / 3$.

In this paper, we present an algorithm that generates a phase estimation circuit capable of perfectly discriminating between any set of phases with certainty after a single run, given the phases are rational multiples of $\pi$. 
This allows one to design a phase estimation circuit with fewer qubits and unitary gates than QPE in many cases.
Notably, special cases of this are seen in the canonical examples of RI and QPE, where RI has a flat prior across two discrete phases, and QPE has a flat prior across $2^n$ discrete phases.
Similarly to these canonical examples, these generated circuits may be run many times to estimate an unknown, continuous phase between the discrete phases.
This algorithm therefore develops a more general class of phase estimation circuits that we call \titlelong\ (\titleabbrev), of which RI and QPE are special cases.
We demonstrate that \titleabbrev\ circuits have a trade-off between phase measurement precision for higher distinguishability of phases, which we show is an interpolation between the canonical RI and QPE circuit. 
This allows one to tune a phase estimation circuit to be optimized for a specific task.

The article is organized as follows.
In Sec.~\ref{Sec:CGA}, we introduce an algorithm that produces \titleabbrev\ circuits and demonstrate its use for two illustrative examples.
Then in Sec.~\ref{Sec:TradeOff}, we analyze how to compare different \titleabbrev\ circuits with a cost-benefit analysis.
We show relevant calculations for our analysis of \titleabbrev\ circuits in Appendix \ref{sec:optimality}.

\section{\titleabbrev\ Generation Algorithm} \label{Sec:CGA}
We begin by presenting an algorithm that generates \titleabbrev\ circuits.
We consider circuits consisting of a set of $\numqub$ qubits, each prepared in the state $\ket{0}$.
We label the qubit states $\ket{q_j}$ with index $j$, we assume that one can perfectly perform instantaneous, noiseless gates and measurements, and we assume no experimental imperfections. 
The gates used in the \titleabbrev\ algorithm are the Z gate $Z_j = \op{0}{0}_j - \op{1}{1}_j$, powers of the Z gate $Z^p_j = \op{0}{0}_j + e^{i \pi p} \op{1}{1}_j$, the Hadamard gate $H_j = (\op{0}{1}_j + \op{1}{0}_j + Z_j) / \sqrt{2}$, and powers of the controlled Z-gate with target $\ket{q_j}$ controlled by $\ket{q_k}$: $C\!Z^p_{j,k} = \mathbb{I}_j \otimes \op{0}{0}_k + Z^p_j \otimes \op{1}{1}_k$.

The objective of quantum phase estimation is to accurately estimate an unknown quantum phase $\theta$ using minimal resources.
In this work, we assume the phase is encoded by $Z$ rotations, such that $\rotgateletter{}_j = e^{-i \theta Z_j / 2}$.
Note that we consider circuits that apply the phase shift directly onto the control qubits through $U_j$, but our results extend to methods that apply the phase shift using any controlled unitary acting on an ancillary register, as is typically done in QPE~\cite{nielsen_chuang_2010}.

We now present an algorithm that generates a circuit that, given some set of phases, $\thetas$ with $|\thetas| = \numthetas$, which is some subset of $\{ \pi \fractionnum{i} = \frac{\pi x_i}{\den} : 0 \leq x_i < 2 \den, x_i \in \mathbb{Z} \}$ where $\mathbb{Z}$ is the set of integers, allows one to determine the encoded phase $\theta \in \thetas$ with certainty after a single run. 
Here, all $\fractionnum{i}$ are rational and can therefore be rewritten with a common integer denominator $\den$ and a set of numerators $S_0 = \{ x_i \}$.
Starting with $i = 0$, the circuit generating algorithm consists of the sequence enumerated in Algorithm~\ref{CircuitAlgorithm}.
\begin{alg}[h]
\begin{enumerate}
    \item Set $\gcds{i}$ as the greatest common divisor (GCD) of the numerators $S_i$.
    
    \item Set $Q_i$ as $\{y = \frac{x}{\gcds{i}} : x \in S_i \}$. 
    
    \item Set $\adds{i}$ as the mode of the differences $y_e - y_o$ for all even $y_e \in Q_i \cap 2 \mathbb{Z}$ and odd $y_o \in Q_i \cap (2 \mathbb{Z} + 1)$ integers.
    If there are no even integers, $Q_i \cap 2 \mathbb{Z} = \varnothing$, $A_i$ is the minimum of $Q_i$.
    
    \item Set $S_{i + 1}$ equal to $Q_i$ with $\adds{i}$ added to the resulting odd integers: $S_{i + 1} = \{ y_e : \forall y_e \in Q_i \cap 2 \mathbb{Z} \} \cup \{ y_o + A_i : \forall y_o \in Q_i \cap (2 \mathbb{Z} + 1) \}$.
    
    \item Iterate (S1)-(S4) by incrementing $i$ until $S_{i + 1} = \{ 0 \}$.

\end{enumerate}

    \caption{Steps of the circuit generating algorithm.
    Note that the set theory notation ignores repeated elements.}
    \label{CircuitAlgorithm}
\end{alg}

To construct the circuit that distinguishes phases in $\thetas$ with certainty after a single run, the gate sequence (see Appendix~\ref{sec:gates})
\begin{equation} \label{eq:AppliedGates}
    H_j \left( \prod_{k = 0}^{j - 1} {C\!Z}_{j,k}^{\frac{A_k}{\prod_{\ell = k + 1}^j G_\ell}} \right) U_j^{\frac{d}{\prod_{\ell = 0}^j G_\ell}} H_j \ket{q_j},
\end{equation}
is applied on each $\ket{q_j}$.
Here, the application of the $H$ and $C\!Z$ gates across all qubits after the unitary applications is reminiscent of the $QFT^\dagger$ gate (see Appendix~\ref{sec:gpe_gate}), $QFT$ standing for the quantum Fourier transform, which can be found in Ref.~\cite{nielsen_chuang_2010}.
Performing a measurement on $\ket{q_j}$ produces a measurement outcome $m_j \in \{0,1\}$, and the measurement of all $n$ qubits produces a binary string $m_0\ldots m_{\highqubind}$. 
This binary string gives an estimation of $\theta$ (see Appendix~\ref{sec:estimate}):
\begin{equation}
    \theta_{\mathrm{est}} = - \frac{\pi}{\den} \sum_{j = 0}^{\highqubind} \measres{j} \adds{j} \prod_{k = 0}^j \gcds{k}.
\end{equation}
We note that \titleabbrev\ circuits can be run sequentially on a single qubit that is reset to $\ket{0}$ after performing each line of the circuit, and each measurement effectively modifies the encoded phase in subsequent unitary applications.
The sequential approach is explored in more depth in Refs.~\cite{mosca1999hidden, Dob_ek_2007} and is effectively equivalent to the multi-qubit, parallel circuits presented here.
In Appendix~\ref{sec:complexity} we analyze the time and space complexity of the generated circuits from Eq.~\eqref{eq:AppliedGates} as well as Algorithm~\ref{CircuitAlgorithm}.
Here, we find that the circuit generating algorithm runs in polynomial time with respect to $\numthetas$ and $\log_2 \hightheta$, where $\hightheta = \max[S_0]$, and the number of qubits needed is upper bounded by $O(\min[\numthetas, \log_2 \hightheta])$.

\begin{figure}
    \centerline{\includegraphics[width=\linewidth]{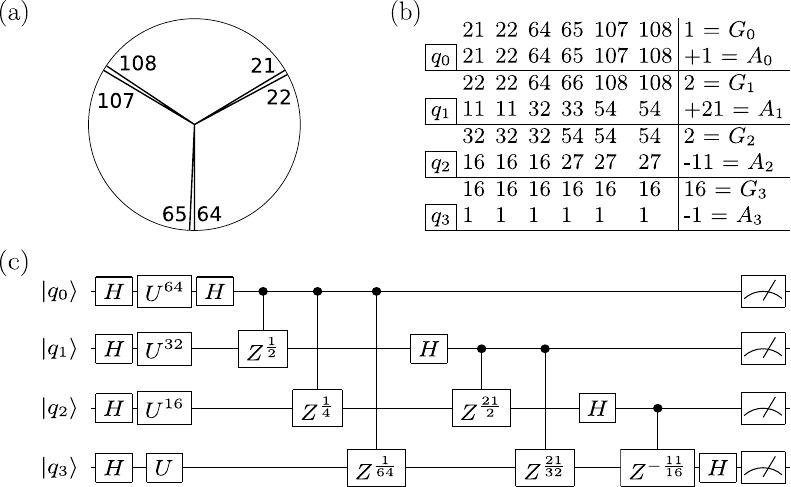}}
    \caption{(a) Visualization of $\thetas = \{ \frac{\pi x}{64} : x \in \{ 21, 22, 64, 65, 107, 108 \} \}$ around the equator of the Bloch sphere. (b) Application of Algorithm~\ref{CircuitAlgorithm} to find $\gcds{i}$ and $\adds{i}$ for all iterations. (c) The circuit generated by Algorithm~\ref{CircuitAlgorithm} by way of Eq.~\eqref{eq:AppliedGates}.}
    \label{fig:gpe_ex_1}
\end{figure}

A common situation in which the \titleabbrev\ circuits of Eq.~\eqref{eq:AppliedGates} can be very useful for hypothesis testing is systems where an external perturbation splits degenerate energy levels.
As an example, we consider a system in which $\ket{1}$ is physically a $F = 1$ hyperfine state while $\ket{0}$ is $F' = 0$. 
Therefore, a magnetic field will cause a Zeeman energy shift of the $\ket{F = 1, m_F = \pm 1}$ states by $\pm \hbar \delta_B$.
Each qubit thus undergoes an additional rotation $\exp[i m_F \delta_B Z_i t / 2]$ for a run time $t$, such that $\theta = \theta' + m_F \delta_B t$.
We therefore may wish to distinguish between three locations on the Bloch sphere's equator to determine which transition is being driven, as shown in Fig.~\ref{fig:gpe_ex_1}(a). 
We thus use Algorithm~\ref{CircuitAlgorithm} to create a circuit that can distinguish the set $\thetas = \{ \frac{\pi x_i}{64} : x_i \in \{21, 22, 64, 65, 107, 108\} \}$ with certainty after a single run. 
In Fig.~\ref{fig:gpe_ex_1}(b), we show the outcome of the iterations of Algorithm~\ref{CircuitAlgorithm} for this particular set of phases.

In the first iteration, $i = 0$, Algorithm~\ref{CircuitAlgorithm} Steps (S1) and (S2) ensures some odd numerators, enabling $\ket{q_0}$ to distinguish between all $\theta_{0,e}$ from all $\theta_{0,o}$ within the set $\{ \frac{\pi \gcds{0} x}{64} : 0 \leq x < \frac{128}{G_0}, x \in \mathbb{Z} \}$. Here, $j$ in $\theta_{i,j}$ indicates whether $x$ is even or odd so that all $\theta_{i,e}$ measure 0 with certainty while all $\theta_{i,o}$ measure 1 with certainty on $\ket{q_i}$.
This is accomplished by the unitary rotation $\rotgateletter{d / \gcds{0}}_0$ as it becomes $\exp[-i x_i \pi Z_j / (2 \gcds{0})]$ with $x_i / \gcds{0} \in \mathbb{Z}$.
The goal of Step (S3) is to pick an addition $\adds{0}$ such that it reduces the size of the set of integers as much as possible, i.e., it converts the maximum amount of odd integers to even integers in the set.
This addition dictates the controlled rotation on subsequent qubits.
Step (S4) then returns a new set of integers $S_1$ from which one finds a new GCD, $\gcds{1}$, in the next iteration. 
This allows one to distinguish between all $\theta_{1,e}$ from all $\theta_{1,o}$ within the set $\{ \frac{\pi \gcds{1} \gcds{0} x}{64} : 0 \leq x < \frac{128}{\gcds{1} \gcds{0}}, x \in \mathbb{Z} \}$.
Step (S5) iterates this process such that each iteration corresponds to a qubit in the generated circuit.

Using this reduction process in conjunction with Eq.~\eqref{CircuitAlgorithm}, we produce the circuit displayed in Fig.~\ref{fig:gpe_ex_1}(c) to distinguish the phases in $\thetas$, where the final symbol stands for an individual qubit measurement in the $Z$-basis.
Note that only four qubits are used to distinguish the desired phases in \titleabbrev\ with certainty after a single run.
This can be compared to seven qubits needed in QPE, since $2 d = 2^7$, demonstrating the utility of our circuit generation algorithm.
Furthermore, while running it once can determine which $\theta \in \thetas$ is encoded, running it many times and using some statistical analysis such as Bayes theorem~\cite{Holland,Reilly3,Wiebe} reliably estimates any continuous $\theta$ within the desired ranges.

\begin{figure}
    \centerline{\includegraphics[width=\linewidth]{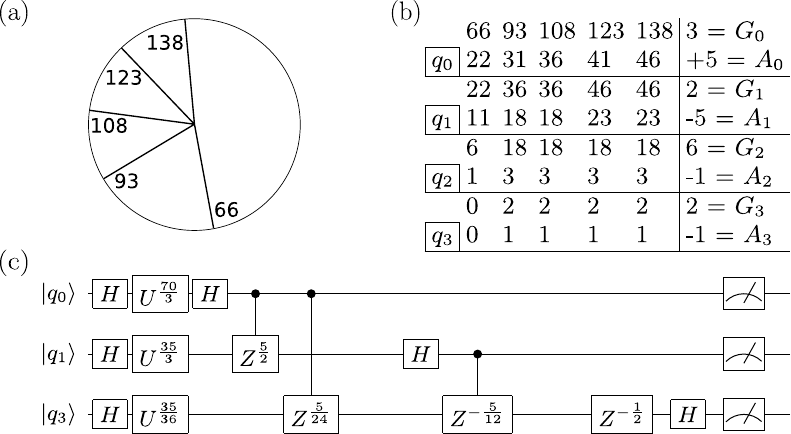}}
    \caption{(a) Visualization of $\thetas = \{ \frac{\pi x}{\exampleden} : x \in \examplethetanumerators \}$ around the equator of the Bloch sphere. (b) Application of Algorithm~\ref{CircuitAlgorithm} to find $\gcds{i}$ and $\adds{i}$ for all iterations. (c) The circuit generated by Algorithm~\ref{CircuitAlgorithm} by way of Eq.~\eqref{eq:AppliedGates} and removing phantom qubits.}
    \label{fig:gpe_ex_2}
\end{figure}

Another interesting feature of Algorithm~\ref{CircuitAlgorithm} can be demonstrated with the example shown in Fig.~\ref{fig:gpe_ex_2}.
Here, we use the circuit generation algorithm to distinguish the phases $\thetas \in \{ \frac{\pi x_i}{70} : x_i \in \{66, 93, 108, 123, 138\} \}$ with certainty after a single run.
As shown in Fig.~\ref{fig:gpe_ex_2}(b), we find a set of only odd numerators in the line for $\ket{\qubnum{2}}$. 
This ensures that, for all phases in $\thetas$, the qubit $\ket{\qubnum{2}}$ will always be in the $\ket{1}$ state when the measurement is performed. 
Therefore, this qubit can be removed from the circuit while $C\!Z_{j,2}^p$ gates can be replaced with uncontrolled $Z_j^p$ gates.
When estimating the original theta, consider this qubit as if it had existed and measured 1, i.e. $m_2 = 1$.
We label $\ket{\qubnum{2}}$ as a ``phantom'' qubit and display the reduced circuit where we have removed the phantom qubit $\ket{q_2}$ in Fig.~\ref{fig:gpe_ex_2}(c).
One can see that the last qubit in Fig.~\ref{fig:gpe_ex_1} is a phantom qubit as well.

Removing a phantom qubit from a circuit does not decrease distinguishability of the discrete phases in $\thetas$ and does not necessarily decrease distinguishability of continuous phase estimation.
We see, for example, that it does not affect distinguishability within the desired ranges in Fig.~\ref{fig:gpe_ex_1}.
Precision, on the other hand, will be affected due to Eq.~\eqref{eq:CFI}.

\section{Trade-off Between Precision and Distinguishability} \label{Sec:TradeOff}
\begin{figure}
    \centerline{\includegraphics[width=\linewidth]{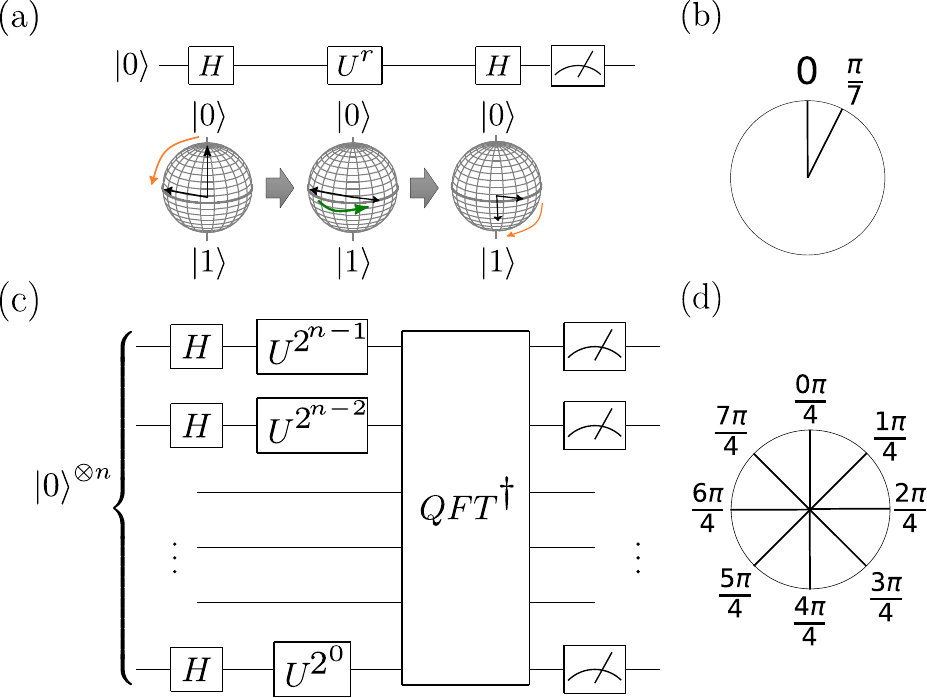}}
    \caption{Canonical phase estimation procedures. (a) RI circuit with the corresponding Bloch sphere rotations. (b) The perfectly distinguishable phases of RI with 7 unitary applications. (c) QPE circuit. (d) The perfectly distinguishable phases of QPE with 3 qubits.}
    \label{fig:current_solutions}
\end{figure}

Interestingly, the extremes of the RQPE circuit generation algorithm create circuits for either RI and QPE, and these serve as perfect examples to showcase these comparable properties.
RI will be automatically generated from Algorithm~\ref{CircuitAlgorithm} when $|\thetas| = 2$, the smallest possible size, while QPE will be automatically generated when $|\thetas| = 2 \den$, the largest possible size.
We display the circuit diagrams for these procedures in Figs.~\ref{fig:current_solutions}(a) and~\ref{fig:current_solutions}(c).
This motivates the comparison of the two extremes, RI and QPE, and the general RQPE algorithm.

\begin{figure}
    \centerline{\includegraphics[width=\linewidth]{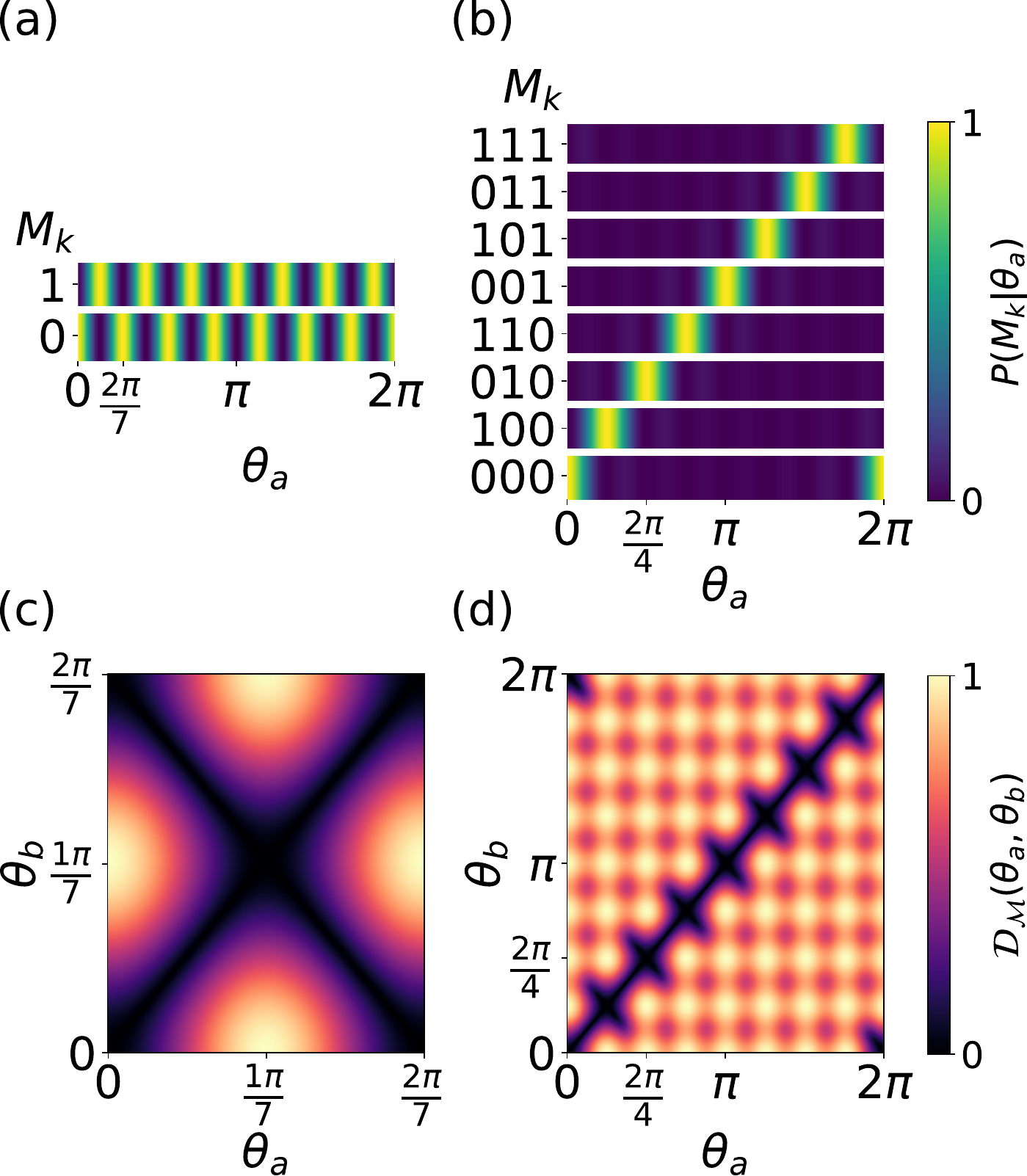}}
    \caption{The conditional probability of (a) RI with 7 unitary applications and (b) QPE with 3 qubits. The distance between phases $\theta_a$ and $\theta_b$ calculated by Eq.~\eqref{eq:distance} for (c) RI with 7 unitary applications and (d) QPE with 3 qubits.}
    \label{fig:probabilities} 
\end{figure}

\titleabbrev\ circuits can be compared by three key properties: the number of unitary applications, distinguishability, and precision.
We use the total number of unitary applications, $\numuni$, during a single run as the constrained resource in order to make a general comparison.
This can be written as a sum over all qubits, $\numuni = \sum_{j=0}^{\highqubind} \uniqub{j}$, where $\uniqub{j}$ represents the number of applications of $\rotgateletter{}_j$ to $\ket{q_j}$. 
In this way, QPE has $\numuni_{\text{QPE}} = 2^\numqub - 1$, and RI has $\numuni_{\text{RI}} = \uniqub{0}$.
One way to compare an RI circuit to QPE would then be to set $\numuni_{\text{RI}} = \numuni_{\text{QPE}}$, thereby applying $(\rotgateletter{}_0)^{\numuni_{\text{QPE}}}$ on the qubit in RI, and compare the resulting precision and distinguishability.
We consider an optimal circuit to be one that minimizes the number of unitary applications while achieving some desired precision and distinguishability, as discussed in more detail in Appendix~\ref{sec:optimality}.

Let $\measoutcome{k}$ be the measured binary string corresponding to the $n$-qubit measurement outcome in the measurement basis $\mathcal{M} = \{ \measoutcome{k} : k \in \mathbb{Z}, 0 \leq k < 2^n \}$.
If, for each $\theta_i \in \thetas$, there exists a unique measurement outcome $\measoutcome{k}$ with conditional probability $P(\measoutcome{k} \lvert \theta_i) = 1$, and for all $\theta_j\in\thetas$ with $\theta_j\neq\theta_i$ the same measurement outcome $\measoutcome{k}$ has conditional probability $P(\measoutcome{k} \lvert \theta_j)=0$, then we say the circuit perfectly distinguishes the set of phases $\thetas$, i.e., distinguishes them with certainty after a single run of the circuit.
Conditional probability, here, is the likelihood of each measurement outcome given a phase~\cite{MandelWolf}.

In general, every phase estimation circuit has a unique set of phases which it can perfectly distinguish.
In RI, one can perfectly distinguish exactly $\numthetas = 2$ phases, $\thetas = \{ 0, \frac{\pi}{r_\mathrm{RI}} \}$, as these are the points where the Ramsey fringes reach their extremum values (see Fig.~\ref{fig:probabilities}(a)). 
For the example circuits we consider, this corresponds to $\theta \in \{ 0, \frac{\pi}{7} \}$ on the equator of the Bloch sphere as shown in Fig.~\ref{fig:current_solutions}(b).
Conversely, the QPE algorithm can perfectly distinguish the phases $\thetas = \{ \frac{\pi x}{2^{\numqub - 1}} : x \in \mathbb{Z}, 0 \leq x < 2^\numqub \}$, such that $\numthetas = 2^\numqub$.
This feature of QPE can be seen in Fig.~\ref{fig:probabilities}(b), where one has $\numthetas=8$ perfectly distinguishable phases corresponding to the eight possible measurement outcomes for $\numqub=3$ qubits.
These perfectly distinguishable phases are shown on the equator of the Bloch sphere in Fig.~\ref{fig:current_solutions}(d).

Using quantum state geometry (adapted from Eq.~(1.57) in~\cite{bengtsson}), one can define the distance between two phases, $\theta_a$ and $\theta_b$, using the $l_2$-distance between conditional probabilities in a measurement basis $\mathcal{M}$:
\begin{equation} \label{eq:distance}
    \mathcal{D}_{\mathcal{M}} (\theta_a, \theta_b) = \sqrt{\frac{1}{2} \sum_{k = 0}^{2^\numqub - 1} \left[ P(M_k \lvert \theta_a) - P(M_k \lvert \theta_b) \right]^2}.
\end{equation}
When $\mathcal{D}_{\mathcal{M}} (\theta_a, \theta_b) = 1$, this is the maximum distance corresponding to two perfectly distinguishable phases, whereas $\mathcal{D}_{\mathcal{M}} (\theta_a, \theta_b) = 0$ corresponds to two phases which cannot be distinguished.
In this way, distinguishability denotes the amount of overlap between probability distributions of two phases.
In Figs.~\ref{fig:probabilities}(c) and~\ref{fig:probabilities}(d), we compare RI and QPE using the distance metric in Eq.~\eqref{eq:distance} and measurements in the $Z$-basis.

While QPE has a larger range of distinguishable phases than RI, this is not the only figure of merit that one wishes to optimize when performing phase estimation.
In the context of quantum metrology, one performs many runs of the circuit to measure $\theta$ from a continuous set of phases, $\thetas_{c} = \{ x \in \mathbb{R}, 0 \leq x < 2 \pi \}$, with greater and greater accuracy (see Fig.~\ref{fig:estimation_graphs}).
Therefore, another important metric of phase estimation circuits is the precision.
This is nicely encapsulated by the classical Fisher information (CFI)~\cite{lehmann_casella_2007} with a given measurement basis $\cfi(\theta|\mathcal{M})$, as the maximal achievable precision over $R$ runs of the circuit is given by the Cram\'er-Rao bound~\cite{Rao} 
\begin{equation} \label{eq:CRB}
    \Delta \theta^2 = \frac{1}{\sqrt{R} \sqrt{ \cfi(\theta|\mathcal{M})}}.
\end{equation}
For the circuits we consider, the CFI for the $Z$-basis is given by (see Appendix~\ref{sec:fi})
\begin{equation} \label{eq:CFI}
    \cfi(\theta|\mathcal{M}) = \sum_{j = 0}^{\highqubind} \uniqub{j}^2.
\end{equation}
One can see that the CFI is dominated by the qubit that has the largest number of unitary applications.
This $\unimax$ in QPE is only on the order of half of $\unimax$ in RI when $\numuni_{\mathrm{RI}} = \numuni_{\mathrm{QPE}}$ because RI has all of its unitary applications acting on a single qubit.
Therefore, QPE will have on the order of half as much precision given the same number of unitary applications.
This can be seen in Figs.~\ref{fig:probabilities}(a) and~\ref{fig:probabilities}(b) by the width of the fringes.

In a general \titleabbrev\ circuit with $\numqub$ qubits, one can perfectly distinguish $\numthetas \leq 2^\numqub$ phases. 
However, the phases need not be evenly distributed around the equator of the Bloch sphere, as is the case with QPE, since the exponential on the unitary gate applications over subsequent qubits are not restricted to powers of two.
Therefore, there is a trade-off between distinguishability and precision that can be tuned for a given parameter estimation objective by employing different \titleabbrev\ circuits.
As with the canonical examples, \titleabbrev\ circuits can either be run once for perfect distinguishability between the phases in $\thetas$ or can be run multiple times to estimate a continuous phase.

\begin{figure}
    \centerline{\includegraphics[width=\linewidth]{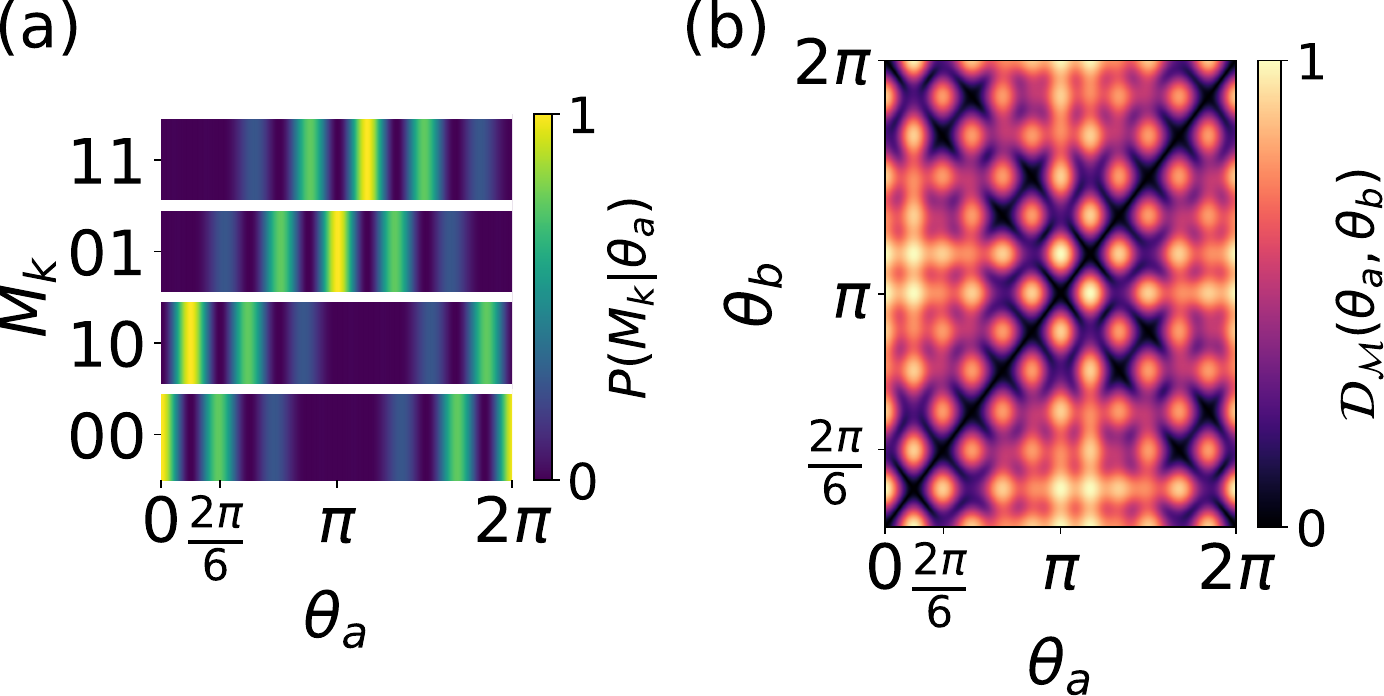}}
    \caption{Using an \titleabbrev\ circuit with 1 and 6 unitary applications on two qubits, shown are (a) the conditional probability and (b) the distance calculated by Eq.~\eqref{eq:distance} between two phases.}
    \label{fig:distances}
\end{figure}
We consider an example circuit in Fig.~\ref{fig:distances}, which is a two-qubit circuit with $\ket{q_0}$ and $\ket{q_1}$.
We use the unitaries $U_0^6$ and $U_1$ to match the number of unitary applications used in the canonical circuits studied in Fig.~\ref{fig:current_solutions}.
The probability distribution for different phases is displayed in Fig.~\ref{fig:distances}(a).
Since the CFI is dominated by $\unimax$, we expect this \titleabbrev\ circuit, having $\unimax = 6$, to have a higher precision than QPE with $\unimax = 4$, but lower than RI with $\unimax = 7$.
This is confirmed in Fig.~\ref{fig:distances}(a) where we compare the width of the first fringe to that of the canonical circuits.

Meanwhile, the opposite relationship is true when comparing the circuit's distinguishablility.
We show this in Fig.~\ref{fig:distances}(b) where we calculate Eq.~\eqref{eq:distance} for our \titleabbrev\ circuit. 
We say that a circuit is more distinguishable if it has a greater number of $\theta_i$ having a distance of 0 to only itself rather than additionally having a distance of 0 to some other $\theta_j$.
Then, this \titleabbrev\ circuit can be compared to Figs.~\ref{fig:probabilities}(a) and~\ref{fig:probabilities}(b) for the canonical examples to see that it is more distinguishable than RI but less distinguishable than QPE. 

There are two notable features of distinguishability in \titleabbrev\ circuits, further analyzed in Appendix~\ref{sec:indistinguishability}. 
One is the repetition of probability distributions. 
This is determined by the qubit with the lowest number of unitary applications causing the probability distributions over the range $\left[ 0, \frac{2 \pi}{\min[u_j]} \right)$ to repeat over the full $2\pi$ range, being truncated after $2\pi$. 
The second distinguishability feature is the distinguishability within this repeated range, which is determined by the distribution of unitary applications over the qubits. The generation algorithm that we have presented in this paper utilizes these two features with the goal of optimizing the distribution of unitary applications for any given $\Theta$.

\begin{figure*}
    \centering
    \includegraphics[width=\linewidth]{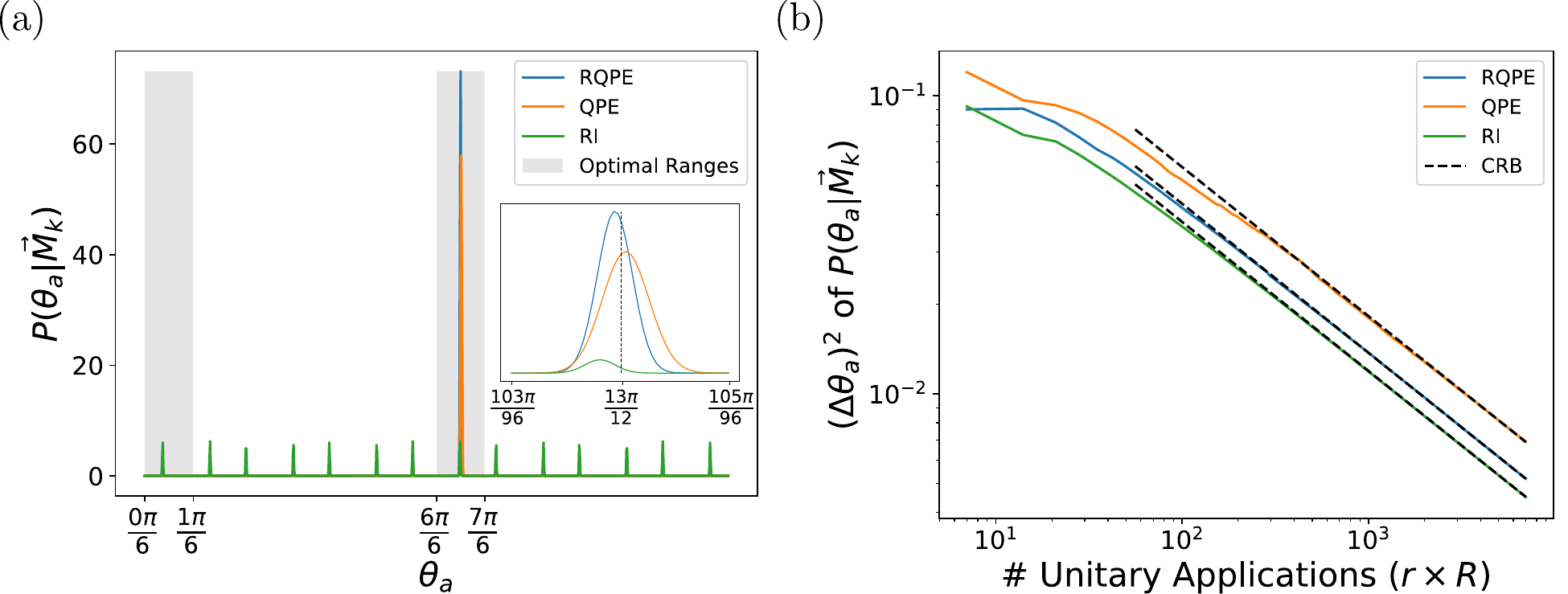}
    \caption{The results of Bayesian reconstruction, Eq.~\eqref{eq:Bayes}, for the three circuits used in Figs.~\ref{fig:probabilities} and~\ref{fig:distances}.
    Here, we choose the actual unknown phase to be $\theta = 13 \pi / 12$ and run each circuit $R_{\mathrm{max}} = 1000$ times.
    (a) The final posterior distribution $P(\theta_a \lvert \vec{M}_k)$ from Eq.~\eqref{eq:Bayes} using an initial flat prior across the range $[0, 2 \pi)$. 
    We also display the ranges of optimal RQPE distinguishability (i.e., regions without aliasing issues) in the shaded gray regions.
    The inset zooms in on the region near $\theta$, which we show as a dotted black line.
    (b) The the standard deviation, $(\Delta \theta_a)^2$, of the posterior distribution $P(\theta_a \lvert \vec{M}_k)$ after $R$ runs of the circuit with an initial flat prior across the range $[\pi, 7\pi / 6]$.
    In general, RQPE circuits should be compared by the number of unitary applications $r \cross R$; here, we have $r = 7$ for all three circuits. 
    We show that for all three circuits in the large run limit $R \gg 1$, the standard deviation converges to the respective Cram\'er-Rao bound (CRB) [dashed black lines], given by Eq.~\eqref{eq:CRB}.}
    \label{fig:estimation_graphs}
\end{figure*}
To demonstrate these properties when estimating an unknown continuous phase, we now implement Bayesian reconstruction~\cite{Holland,Reilly3} of a non-perfectly distinguishable phase $\theta = 13 \pi / 12 \notin \Theta$ in Fig.~\ref{fig:estimation_graphs} for the three circuits considered in Figs.~\ref{fig:probabilities} and~\ref{fig:distances}. 
This takes into account non-deterministic wavefunction collapse, and closer represents traditional phase estimation used by, for example, atomic clocks.
Here, we run the complete circuit $R$ times which produces a measurement record $\vec{M}_k = \left[ M_k^{(1)}, M_k^{(2)}, \ldots, M_k^{(R)} \right]$, where $M_k^{(j)}$ is the total measurement outcome of the $j$th run of the circuit.
We then iterate Bayes theorem to find 
\begin{equation} \label{eq:Bayes}
    P(\theta_a \lvert \vec{M}_k) = P(\theta_a) \prod_{j = 1}^{R} \frac{P(M_k^{(j)} \lvert \theta_a)}{P(M_k^{(j)})},
\end{equation}
where $P(M_k^{(j)} \lvert \theta_a)$ is from the forward problem [i.e., the rows of Figs.~\ref{fig:probabilities}(a),~\ref{fig:probabilities}(b), and~\ref{fig:distances}(a)], $P(\theta_a)$ is the prior knowledge, and $P(M_k^{(j)})$ is a normalization factor.
The results for all three circuits are shown with $R_{\mathrm{max}} = 1000$ in Fig.~\ref{fig:estimation_graphs}(a), where we choose the prior as a flat distribution. 
The posterior distribution $P(\theta_a \lvert \vec{M}_k)$ of the three schemes show that the peak of \titleabbrev\ is higher than QPE and RI, signifying greater confidence in estimates of  $\theta$. 
Both RQPE and QPE do not suffer from the aliasing problems that RI has for this value of $\theta$ (i.e., the many green peaks).
The inset confirms that the posterior distribution is converging to the correct value of $\theta$ (dotted black line) for \titleabbrev.
We also display the ranges where the \titleabbrev\ circuit does not have aliasing problems in the shaded gray regions of Fig.~\ref{fig:estimation_graphs}(a), and so for a $\theta$ in the non-shaded regions, one would expect \titleabbrev\ to have $2$ peaks in the two non-shaded regions due to indistinguishability (see Fig.~\ref{fig:distances}(b)).
Note that this gray region corresponds to the prior that generates the \titleabbrev\ circuit corresponding to Fig.~\ref{fig:distances}.

To compare the sensitivity of the three circuits, we choose a flat prior between $[\pi, 7 \pi / 6]$ and again iterate Bayes theorem Eq.~\eqref{eq:Bayes} for $R_{\mathrm{max}} = 1000$ runs in Fig.~\ref{fig:estimation_graphs}(b). 
We calculate the standard deviation of the posterior distribution $(\Delta \theta_a)^2$ after each run of the circuits which we show on a log-log plot. 
To keep the three circuits on the same footing, we use the total number of unitary applications $r \cross R$ rather than just the number of runs of the circuit $R$ on the x-axis.
Here, this distinction is superfluous as we have chosen $r_{\mathrm{RI}} = r_{\mathrm{QPE}} = r_{\mathrm{\titleabbrev}} = 7$, but this is not true for general \titleabbrev\ circuits. 
We display the Cram\'er-Rao bound Eq.~\eqref{eq:CRB} as a dashed black line for each respective circuit, and one can see that all three circuits converge to the Cram\'er-Rao bound in the large run limit $R \gg 1$.
This would not be true for RI with a flat prior across the whole $[0, 2 \pi)$ range, as was done in Fig.~\ref{fig:estimation_graphs}(a), unless one fit the posterior distribution to multiple Gaussian functions. 
We also note that the standard deviation converges from below the Cram\'er-Rao bound in all three case due to our choice in prior, which has a width initially below the Cram\'er-Rao bound.
In all three cases, we asymptotically converge to the Cram\'er-Rao bound and match the analysis of the different CFIs.

\section{Conclusion and outlook} \label{Sec:Conclusion}
In this article, we have demonstrated that a general class of quantum phase estimation algorithms, which we labeled as \titleabbrev, exist that encompass the canonical examples of RI and QPE.
Furthermore, by casting these canonical examples into the language of \titleabbrev\, we are able to compare these distinct algorithms on equal footing.
The figures of merit that we considered were precision, determined by the CFI in Eq.~\eqref{eq:CFI}, and the distinguishablility, determined by the distance in Eq.~\eqref{eq:distance}.
The figures of cost that we considered were the number of unitary applications and the number of qubits.
We found that RI is more sensitive than QPE when constrained to the same number of unitary applications, but QPE is more distinguishable than RI.
We demonstrated that these are two extremes of \titleabbrev\ circuits, and so one can find a middle ground between these examples by tuning a trade-off between these two figures of merit.

While the presented circuit generation algorithm can be used to improve current phase estimation standards, it is not necessarily optimal in all cases.
Future work may therefore be directed towards either reducing complexity in the generation algorithm or increasing optimality in the generated circuits.
Future work will also explore how to best implement RQPE in contexts beyond quantum computing circuits, such as quantum optics systems that have, so far, primarily utilized RI for phase estimation~\cite{Huang,ActuallyJarrodsPaper,Reilly2}. 
Of particular interest, these systems will offer the opportunity to study the interplay of experimental errors with the algorithm, and the opportunity to benefit from error correction.

\section*{Acknowledgments}
The authors thank J. Cooper, S. B. Jäger, and B. Waggoner for useful discussions. 
This material is based upon work supported in part by the U.S. Department of Energy, Office of Science, National Quantum Information Science Research Centers, Quantum Systems Accelerator. 
This research was also supported by NSF OSI Grant No. 2231377, NSF PHY Grant No. 2317149, NSF OMA Grant No. 2016244, and NSF PHY Grant No. 2207963.

\appendix

\section{Gates on a Qubit}\label{sec:gates}

In this section, we derive the formula to calculate the gates applied to each qubit based on the reductions in Algorithm~\ref{CircuitAlgorithm}.
First, we assume the phase can be written as $\theta = \frac{\pi x_a}{\den}$ for integers $x_a$ and $d$. 
The goal is to perform some operations to change $\theta$ so that the $j$th qubit measures $\rotgate{\pi x_j}$, where $x_j$ is the numerator from the set $\mathcal{F}_j = \{ \frac{x}{\gcds{j}} : x \in \numerators{j} \}$ to which $x_a$ transforms. 
For $x_0$, the algorithm only needs to divide by $\gcds{0}$, resulting in the following gate 
\begin{equation}
\begin{aligned}
    \rotgate{\pi x_0} &= \rotgate{\pi \frac{x_a}{\gcds{0}}} \\
    &= \rotgate{\pi \frac{x_a \den}{\gcds{0} \den}} \\
    &= \left( \rotgate{\pi \frac{x_a}{\den}} \right)^\frac{\den}{\gcds{0}} \\
    &= \rotgateletter{\frac{\den}{\gcds{0}}}_i.
\end{aligned}
\end{equation}
Now doing the same process for $x_1$, the previous reduction applies as well as $\adds{0}$ and $\gcds{1}$, adding $\adds{0}$ only if the numerator at $\mathcal{F}_0$ is odd, which happens when $\qubnum{0} = 1$.
This results in
\begin{equation}
\begin{aligned}
    x_1 &= \left( \frac{x_a}{\gcds{0}} + \qubnum{0} \adds{0} \right)\frac{1}{\gcds{1}} \\
    &= \frac{x_a}{\gcds{0} \gcds{1}} + \frac{\qubnum{0} \adds{0}}{\gcds{1}} \\
    &= \frac{x_a \den}{\gcds{0} \gcds{1} \den} + \frac{\qubnum{0} \adds{0}}{\gcds{1}} \\
    &= \frac{x_a}{\den} \frac{\den}{\gcds{0} \gcds{1}} + \frac{\qubnum{0} \adds{0}}{\gcds{1}},
\end{aligned}
\end{equation}
and so
\begin{equation}
\begin{aligned}
    e^{- i \pi x_1 Z / 2} &= 
    \exp\left[- i \pi \left(\frac{x_a}{\den} \frac{\den}{\gcds{0} \gcds{1}} + \frac{\qubnum{0} \adds{0}}{\gcds{1}}\right) \frac{Z}{2} \right] \\
    &= \left(e^{-i \pi \frac{x_a}{d} \frac{Z}{2}}\right)^\frac{\den}{\gcds{0} \gcds{1}} e^{-i \pi \frac{\qubnum{0} \adds{0}}{\gcds{1}} \frac{Z}{2}} \\
    &= \rotgateletter{\frac{\den}{\gcds{0} \gcds{1}}}_j Z_j^\frac{\qubnum{0} \adds{0}}{\gcds{1}} \\
    &= Z_j^\frac{\qubnum{0} \adds{0}}{\gcds{1}} \rotgateletter{\frac{\den}{\gcds{0} \gcds{1}}}_j.
\end{aligned}
\end{equation}
One can thus see that this pattern results in the following gates for qubit $\qubnum{j}$:
\begin{equation}
\begin{aligned}
     &H_j \left( Z_j^\frac{\qubnum{0} \adds{0}}{\gcds{1} \ldots \gcds{i}} Z_j^\frac{\qubnum{1} \adds{1}}{\gcds{2} \ldots \gcds{j}} \ldots Z_j^\frac{\qubnum{j-1} \adds{j-1}}{\gcds{j}} \right) \rotgateletter{\frac{\den}{\gcds{0} \ldots \gcds{j}}}_j H_j \\
    = &H_j \left( \prod_{k = 0}^{j - 1} C\!Z_{j,k}^{\frac{A_k}{\prod_{\ell = k + 1}^j G_\ell}} \right) U_j^{\frac{d}{\prod_{\ell = 0}^j G_\ell}} H_j.
\end{aligned}
\end{equation}

\section{Estimating \texorpdfstring{$\theta$}{θ}}\label{sec:estimate}

In this section, we derive the formula to estimate the phase from the measurement results after running the circuit generated from Algorithm~\ref{CircuitAlgorithm}.
Backtracking through the reductions (i.e., subtracting the additions and multiplying by the GCDs) reveals the value assigned to each qubit measurement, which we call the bit value. 
For example, in a four-qubit circuit, the original numerator will be
\begin{equation}
\begin{aligned}
    \theta &= \pi ((((-\measres{3} \adds{3}) \gcds{3} - \measres{2} \adds{2}) \gcds{2} \\
    &\quad - \measres{1} \adds{1}) \gcds{1} - \measres{0} \adds{0}) \gcds{0}\\
    &= \pi (-\measres{3} \adds{3} \gcds{3} \gcds{2} \gcds{1} \gcds{0} - \measres{2} \adds{2} \gcds{2} \gcds{1} \gcds{0} \\
    &\quad - \measres{1} \adds{1} \gcds{1} \gcds{0} - \measres{0} \adds{0} \gcds{0}).
\end{aligned}
\end{equation}
One can see the pattern that the bit value of each measured bit $b_i$, which is included in the final estimate if and only if $m_j = 1$, will be
\begin{equation}
\begin{aligned}
    b_j &= - \frac{\pi}{d} \adds{j} \prod_{k=0}^j \gcds{k}.
\end{aligned}
\end{equation}
This allows for the transformation of the measurement results $\measres{0}, \dots, \measres{\highqubind}$ into the unknown phase as
\begin{equation}
\begin{aligned}
    \theta &= -\frac{\pi}{\den} \sum_{j=0}^{\highqubind} \measres{j} \adds{j} \prod_{k=0}^j \gcds{k}. \label{eq:estimate}
\end{aligned}
\end{equation}

\section{Complexity}\label{sec:complexity}

In this section, analyze the time and space complexity of Algorithm~\ref{CircuitAlgorithm} and the generated \titleabbrev\ circuits.
We define $\hightheta$ to be the greatest numerator in $\numerators{0}$ and note that $\numthetas \leq 2 \den$ and $\hightheta < 2 \den$.
Each qubit will ideally reduce the size of $\thetas$ by half, imposing a lower bound of $\lceil \log_2 (\numthetas) \rceil$ on the number of qubits, $\numqub$.
Additionally, the algorithm will always be able to reduce at least one $\theta$ into another after each measurement, and the final measurement distinguishes two distinct values of $\theta$, and so $\numqub \leq \numthetas - 1$.

Moreover, any set of phase numerators is a subset of $\{ i : i \in \mathbb{Z} \text{ and } 0 \leq i < 2^{\lfloor \log_2 \hightheta \rfloor + 1} \}$, which can be distinguished using $\adds{} = [-1, -1, -1, \dots]$ and $\gcds{} = [1, 2, 2, 2, \dots]$ with $\lfloor \log_2 \hightheta \rfloor + 1$ qubits.
If one restricts the additions used in the classical generation algorithm to negative values, then the number of qubits will always be less than or equal to this.
However, allowing positive additions further optimizes many circuits, and since we cannot guarantee this constraint in this case, it can be trivially ensured by using these $\adds{}$ and $\gcds{}$ values if the generated circuit uses more qubits than this.
This puts a second upper bound on $\numqub$, resulting in the following bounds:
\begin{equation}
    \lceil \log_2 (\numthetas) \rceil \leq \numqub \leq \min(\numthetas - 1, \lfloor \log_2 \hightheta \rfloor + 1).
    \label{eq:bound_qubits}
\end{equation}
In the worst case, the generation algorithm will produce this $\gcds{}$, resulting in the geometric sequence
\begin{equation}
    a_i = \den \left( \frac{1}{2} \right)^{i}, \label{eq:qubit_unitaries}
\end{equation}
where $a_i$ is the number of unitary applications on $\qubnum{i}$. This series results in an upper bound on the total number of unitary applications, $\numuni$, in one run of the circuit as
\begin{align}
    \numuni &< 2 \den. \label{eq:total_unitaries}
\end{align}

Each reduction in the generation algorithm is composed of first finding the GCD of the set of values of $\theta$.
Finding the GCD of two numbers $a$ and $b$ can be done using the Euclidean GCD Algorithm in $O(\log_2\min(a, b))$ time \cite{bach_shallit_1997}.
The algorithm can then perform this iteratively for all the possible values of $\theta$ in a set, achieving a time complexity of $O(\numthetas \log_2 \hightheta)$.

The second phase of a reduction consists of finding a value to add to the odd numerators which reduces the set size as much as possible.
The method described in this paper runs in at most $O(\numthetas^2)$ time, since in the worst case there will be an equal number of even and odd numerators, $\frac{\numthetas}{2}$, and each odd will be added to each even, producing $\frac{\numthetas^2}{4}$ results.
Each result is then counted to find the mode, which results in a total of $\frac{\numthetas^2}{4}$ time in the worst case.

Each multiplication and addition can be considered to run in constant time (assuming all numbers fit into the word size of the hardware) over each $\theta$, so the operations after finding a GCD or addition can be done in time $O(\numthetas)$.

The combination of all of these steps within an iteration multiplied by the number of iterations necessary results in a total time complexity of $O((\numthetas^2 + \numthetas \log_2 \hightheta) \cdot \min(\numthetas, \log_2 \hightheta))$ to generate the quantum circuit. Let's look at each case more specifically.
In the case that $\numthetas \leq \log_2 \hightheta$:
\begin{equation}
\begin{aligned}
    O((\numthetas^2 + \numthetas \log_2 \hightheta) \cdot \numthetas) &= O(\numthetas^3 + \numthetas^2 \log_2 \hightheta) \\
    &= O(\numthetas^2 (\numthetas + \log_2 \hightheta)) \\
    &= O(\numthetas^2 \log_2 \hightheta),
\end{aligned}
\end{equation}
because $\log_2 \hightheta$ is the larger term in $(\numthetas + \log_2 \hightheta)$.
In the case that $\numthetas \geq \log_2 \hightheta$:
\begin{equation}
\begin{aligned}
    O((\numthetas^2 + \numthetas \log_2 \hightheta) \cdot \log_2 \hightheta) &= O(\numthetas^2 \log_2 \hightheta + \numthetas (\log_2 \hightheta)^2) \\
    &= O(\numthetas \log_2 \hightheta (\numthetas + \log_2 \hightheta)) \\
    &= O(\numthetas^2 \log_2 \hightheta),
\end{aligned}
\end{equation}
because $\numthetas$ is the larger term in $(\numthetas + \log_2 \hightheta)$.
Therefore, both cases end up being the same and produce a runtime of
\begin{equation}
\begin{aligned}
    O(\numthetas^2 \log_2 \hightheta).
\end{aligned}
\end{equation}

The algorithm stores the possible values of $\theta$ and two arrays for the GCDs and additions in each reduction, which are all on the order $O(\numthetas)$.
In this paper's method to find a particular addition in a reduction step, however, the algorithm stores $O(\numthetas^2)$ possibilities to consider, which is the dominant factor.
This equates to a total space complexity of $O(\numthetas^2)$.

\section{Building a Circuit from Bit Values}\label{sec:bit_values}

In this section, we show how to reverse engineer Algorithm~\ref{CircuitAlgorithm}, deriving the $\adds{}$ and $\gcds{}$ starting with bit values.
The goal of the circuit-building reductions is to find the $\adds{}$s and $\gcds{}$s.
If one is given the bit values instead of a $\thetas$ set, thereby distinguishing $\thetas$ consisting of all possible subset sums of these bit values, one can find these values with much more simplicity.
First, one finds $S_0$ and $d$, which can be done by first finding the fraction forms using the continued fractions algorithm~\cite{continuedFractions}.
Then, set $\gcds{0} = \texttt{GCD}(S_0)$ and $\adds{0} = -\frac{S_{0, 0}}{\gcds{0}}$, where $S_{0, j}$ is the $j$th numerator corresponding to $b_j$.
Next, set $\gcds{1} = \texttt{GCD}(\{ \frac{S_{0,j}}{\gcds{0}} : j \geq 1\})$ and $\adds{1} = -\frac{S_{0,1}}{\gcds{0} \gcds{1}}$.
Continue this way until all $\gcds{}$s and $\adds{}$s are found.

Note that all $\adds{}$ must be odd valued and all $\gcds{}$, except $\gcds{0}$, must be even valued for the presented circuit generation to be sure to work correctly.
If these do not hold, then the presented circuit generation algorithm likely cannot be used to distinguish a $\thetas$ set matching these bit values exactly.

For example, if one wanted to build a circuit that distinguishes phases in the range $[0, \frac{21 \pi}{d}]$ in even multiples of $\frac{3 \pi}{d}$, one could use the bit values $b = [\frac{3 \pi}{d}, \frac{6 \pi}{d}, \frac{12 \pi}{d}]$.
\begin{equation}
\begin{aligned}
    \gcds{0} &= \texttt{GCD}(S_0) \\
    &= 3, \\
\end{aligned}
\end{equation}
\begin{equation}
\begin{aligned}
    \adds{0} &= -\frac{S_{0,0}}{\gcds{0}} \\
    &= -\frac{3}{3} \\
    &= -1, \\
\end{aligned}
\end{equation}
\begin{equation}
\begin{aligned}
    \gcds{1} &= \texttt{GCD}(\left\{ \frac{S_{0,1}}{\gcds{0}}, \frac{S_{0,2}}{\gcds{0}} \right\}) \\
    &= \texttt{GCD}(\left\{ \frac{6}{3}, \frac{12}{3} \right\}) \\
    &= \texttt{GCD}(\{2, 4\}) \\
    &= 2, \\
\end{aligned}
\end{equation}
\begin{equation}
\begin{aligned}
    \adds{1} &= -\frac{S_{0,1}}{\gcds{0} \gcds{1}} \\
    &= -\frac{6}{3 \cdot 2} \\
    &= -1, \\
\end{aligned}
\end{equation}
\begin{equation}
\begin{aligned}
    \gcds{2} &= \texttt{GCD}(\left\{ \frac{S_{0,2}}{\gcds{0} \gcds{1}} \right\}) \\
    &= \texttt{GCD}(\left\{ \frac{12}{3 \cdot 2} \right\}) \\
    &= 2, \\
\end{aligned}
\end{equation}
\begin{equation}
\begin{aligned}
    \adds{2} &= -\frac{S_{0,2}}{\gcds{0} \gcds{1} \gcds{2}} \\
    &= -\frac{12}{3 \cdot 2 \cdot 2} \\
    &= -1.
\end{aligned}
\end{equation}
Since only the first $\gcds{}$ is odd and all the $\adds{}$ are odd, this $b$ is valid, and these $\adds{}$s and $\gcds{}$s may be used in the rest of the circuit generation.

\section{Fisher Information}\label{sec:fi}

In this section, we derive the formula to calculate the Fisher information of an \titleabbrev\ circuit.
As explored in~\cite{mosca1999hidden, Dob_ek_2007}, \titleabbrev\ circuits may be equivalently run sequentially on a single qubit, so each line in the circuit will be run separately from each other. This means that each line of the circuit can be considered a Ramsey Interferometry (RI) circuit with additional Z-rotation gates. These rotation gates have the following effect
\begin{equation}
\label{eq:ramseyMatrix}
\begin{aligned}
H(Z^{z_0}Z^{z_1}\dots)U^\numuni H\ket{0} &= H(Z^{z_0 + z_1\dots})U^\numuni H\ket{0} \\
    &= HZ^jU^\numuni H\ket{0} \\
    &= \frac{1}{2}\begin{bmatrix}
        1 + e^{i (\numuni \theta + j)} \\
        1 - e^{i (\numuni \theta + j)}
    \end{bmatrix},
\end{aligned}
\end{equation}
where $z_i$ is the power of the $i$th Z-rotation gate, and $j$ is a simplifying variable $j = \sum_{i=0} z_i$. 

We now let $E[a]$ be the $a$th element of Eq.~\eqref{eq:ramseyMatrix}.
The classical Fisher information of Eq.~\eqref{eq:ramseyMatrix} is then (see Ref.~\cite{Paris})
\begin{equation}
    \begin{aligned}
        \mathcal{I} \left( \theta \lvert{\mathcal{M}} \right) [\mathrm{RI}] &= \sum_{a=0}^1 \frac{\left\{ \frac{\partial}{\partial \theta} (E[a] E[a]^*) \right\}^2}{E[a] E[a]^*} \\
        &= \frac{\left\{ \frac{\partial}{\partial \theta} \left[\cos(\frac{r \theta + j}{2})\right]^2 \right\}^2}{\left\{\cos(\frac{r \theta + j}{2})\right\}^2} + \frac{\left\{ \frac{\partial}{\partial \theta} \left[ \sin(\frac{r \theta + j}{2}) \right]^2 \right\}^2}{\left\{ \sin(\frac{r \theta + j}{2})\right\}^2} \\
        &= \frac{\left\{ -\frac{r}{2} \sin(r \theta + j) \right\}^2}{\left\{ \cos(\frac{r \theta + j}{2}) \right\}^2} + \frac{\left\{ \frac{r}{2} \sin(r \theta + j) \right\}^2}{\left\{ \sin(\frac{r \theta + j}{2}) \right\}^2} \\
        &= \frac{r^2}{4} \left\{ \frac{\left[ \sin(r \theta + j)\right]^2}{\left[ \cos(\frac{r \theta + j}{2})\right]^2} + \frac{\left[ \sin(r \theta + j) \right]^2}{\left[ \sin(\frac{r \theta + j}{2}) \right]^2} \right\} \\
        &= r^2 \left\{ \left[ \sin(\frac{r \theta + j}{2}) \right]^2 + \left[ \cos(\frac{r \theta + j}{2}) \right]^2 \right\} \\
        &= \numuni^2,
    \end{aligned}
\end{equation}
where $\numuni$ is the number of unitary applications on the qubit. 
One can see that the Fisher information of an RI circuit is irrespective of the embedded $\theta$ or $j$. 
Hence, additional rotation gates on a Ramsey line have no effect on the the Fisher information of that line, and we may simply add the Fisher information of all lines in the \titleabbrev\ circuit to calculate the full Fisher information with
\begin{equation}
    \mathcal{I} \left( \theta \lvert\mathcal{M} \right) [\mathrm{\titleabbrev}] = \sum_{i = 0}^{\highqubind} \uniqub{i}^2.
\end{equation}

\section{Examples of Comparing Optimality}\label{sec:optimality}

In this section, we compare RI, QPE, and RQPE circuits in a more general manner to find conditions for which one might be more optimal than the other.
When one implements a phase estimation circuit, one typically desires some precision, defined by the Cram\'er-Rao bound, while being able to distinguish some range of phases $0 \leq \theta \leq h$ in $e^{-i \theta Z / 2}$, where $h$ is the largest phase, after many runs. 
This is because if there were separate ranges, RI would still need to distinguish one large range that covers all the separate ranges.
A more optimal circuit achieves both of these things while using fewer unitary application and qubits. 
Define precision as $\frac{1}{p}$ where $p \geq 1$ and $h = \frac{a}{d}$. 
The precision is defined by the Cram\'er-Rao bound, so $p = \sqrt{R} \sqrt{\mathcal{I} (\theta \lvert \mathcal{M})}$, where $R$ is the total number number of runs and $\mathcal{I}$ is the classical Fisher information.

First, we see what RI requires. 
RI optimally satisfies these requirements with $\numuni = \frac{1}{h} = \frac{d}{a}$ unitary applications per run. 
Increasing this shrinks the dynamic range to be less than the required theta range, and distinguishability is lost, while 
decreasing this exponentially increases the total number of runs to achieve the same precision, resulting in more unitary applications overall. 
Since the classical Fisher information is $(\frac{d}{a})^2$ according to Eq.~\eqref{eq:CFI}, we have
\begin{equation}
\begin{aligned}
    \frac{1}{p} &= \frac{a}{d\sqrt{R}} \\
    t &= \left\lceil \left( \frac{pa}{d} \right)^2 \right\rceil.
\end{aligned}
\end{equation}
Here, we take the ceiling because $R$ is an integer, meaning RI requires $\left\lceil \left( \frac{pa}{d} \right)^2 \right\rceil$ runs to satisfy the requirements. Since each RI run uses exactly one qubit, it requires the same number of qubits. The total number of unitary applications it needs across all runs is therefore 
\begin{equation}
\begin{aligned}
    \frac{d}{a} R &\geq \frac{p^2 a}{d}.
\end{aligned}
\end{equation}

Quantum phase estimation (QPE) always has full distinguishability over $[0, 2 \pi)$ and achieves at least the desired precision when it's most precise qubit reaches it, which happens after 1 run when
\begin{equation}
\begin{aligned}
    \frac{1}{p} &= \frac{1}{\sqrt{\mathcal{I}}} \\
    p &\geq u_{max}
\end{aligned}
\end{equation}
where $u_{max}$ is the number of unitary applications on the most precise qubit. 
Since the unitary applications for each qubit in QPE follow powers of two, the most precise qubit having $2^{\lceil \log_2 p \rceil} \leq 2^{\log_2 (p) + 1}$ unitary applications satisfies the precision requirement. 
This, in turn, results in less than $4p$ total unitary applications across all qubits and $\lceil \log_2 p \rceil + 1$ total qubits.

We now have circuits for both RI and QPE that satisfy the distinguishability requirement, so, in order for QPE to outperform RI, we find a precision where the total number of unitary applications for QPE is less than that of RI:
\begin{equation}
\begin{aligned}
    4p < \frac{p^2 a}{d} \\
    4p < p^2 h \\
    \frac{4}{p} < h.
\end{aligned}
\end{equation}
This means that QPE is guaranteed to use fewer unitary applications whenever the highest $\theta$ is at least four times the desired precision, i.e. when you would like to estimate $\theta$ using at least 4 bins of equal spacing.

We do a similar analysis for \titleabbrev. 
If one wanted to split the range into between some positive integer $k$ and $k + 1$ bins, we have
\begin{equation}
\begin{aligned}
    \frac{1}{p} &= \frac{b}{c}h \\
    &= \frac{ba}{cd},
\end{aligned}
\end{equation}
where $\frac{c}{k + 1} < b < \frac{c}{k}$. 
One could, for example, run Algorithm~\ref{CircuitAlgorithm} with $\thetas = \{ \frac{b a x_i}{cd} : x_i \in \mathbb{Z}, 0 \leq x_i \leq k + 1 \}$, which satisfies the precision and distinguishability requirements after a single run. We will compare the optimality of this circuit, which uses
\begin{equation}
\begin{aligned} 
    r_{RQPE} &= \frac{cd}{ba}(1 + \frac{1}{2} + \frac{1}{4} + \dots + \frac{1}{2^{\lfloor \log_2 k \rfloor}}) \\
    &= \frac{cd}{ba}\left( 2 - \frac{1}{2^{\lfloor \log_2 k \rfloor}}) \right).
\end{aligned}
\end{equation}
Since RI requires 
\begin{equation}
\begin{aligned}
    t &= \left\lceil \left( \frac{pa}{d} \right)^2 \right\rceil \\
    &= \left\lceil \left( \frac{cda}{bad} \right)^2 \right\rceil \\
    &= \left\lceil \left( \frac{c}{b} \right)^2 \right\rceil \\
    &> k^2 \\
    &\geq k^2 + 1
\end{aligned}
\end{equation}
runs to reach the desired precision, RI is more optimal than this \titleabbrev\ circuit when 
\begin{equation}
\begin{aligned}
    \left\lceil \left( \frac{c}{b} \right)^2 \right\rceil \left( \frac{d}{a} \right) &< \frac{cd}{ab} \left( 2 - \frac{1}{2^{\lfloor \log_2 k \rfloor}}) \right) \\
    \left\lceil \left( \frac{c}{b} \right)^2 \right\rceil &< \frac{c}{b} \left( 2 - \frac{1}{2^{\lfloor \log_2 k \rfloor}}) \right) \\
    k^2 + 1 &< \frac{c}{b} \left( 2 - \frac{1}{2^{\lfloor \log_2 k \rfloor}}) \right) \\
    b &< \frac{c}{k^2 + 1} \left( 2 - \frac{1}{2^{\lfloor \log_2 k \rfloor}}) \right). \\
\end{aligned}
\end{equation}
When $k=0$, the given \titleabbrev\ circuit simply returns RI, so they are equally optimal. 
For $k = 1$, we see that RI is more optimal when $b < \frac{3c}{4}$, i.e. one wants a precision of $\frac{3}{4}h$. For $k = 2$, $b < \frac{3c}{10}$, which violates $b > \frac{c}{3}$.
This violation persists for all $k > 1$. 
Therefore, this \titleabbrev\ circuit is always more optimal than RI except for in the very specific case when one wants $\frac{3}{4}h < \frac{1}{p} \leq h$.

\section{Features of Indistinguishability}\label{sec:indistinguishability}

In this section, we explain two key types of distinguishability that can be seen in \titleabbrev\ circuits.
For all \titlelong\ circuits, a single qubit can distinguish between a phase of 0 and $\frac{\pi}{\uniqub{i}}$ with certainty. 
However, it also creates a range $\left[ 0, \dyrange \right)$ where each $\theta$ within it results in an identical Bloch sphere positions as $\theta + j\dyrange : j \in \mathbb{N}, j < \max[1, \uniqub{i}]$, where $\mathbb{N}$ are the natural numbers. 
That is, on the full $2 \pi$ range of possible phases, there will be $\max[1, \uniqub{i}]$ sets of physically indistinguishable phases after the unitary applications. 
We call this range the repeated range, $\mathcal{S}$, and we call this principle of indistinguishable sets ``indistinguishability due to repitition''. 
This can be extended to a multi-qubit \titleabbrev\ circuit, where
\begin{align}
    \mathcal{S} = \dyrange,
\end{align}
as $\unimin$ has the largest repeated range, and no repetition occurs until repetition over this largest repeated range occurs.

Another type of indistinguishability, which we refer to as ``indistinguishability due to measurement basis'', is due to identical measurement probability distributions for phases within the repeated range itself. 
This means that the state may lie on different points in the Bloch sphere after the unitary applications but have identical probability distributions given the measurement basis. 
This is the type of distinguishability that can be increased by distributing unitary applications across multiple qubits in certain ways. 
This is also the type that is plotted in our distance graphs, since this distinguishability is simply repeated as the repeated range repeats.

\section{\titleabbrev\ as a Gate}\label{sec:gpe_gate}

In the same way that the $H$ and $C\!Z$ gates after the unitary applications of a QPE circuit can be viewed as a $QFT^\dagger$ gate, one can view the $H$ and $C\!Z$ gates after the unitary applications of an \titleabbrev\ circuit as a $(\titleabbrev_{\gcds{}, \adds{}})^{\dagger}$ gate. 
In the following definitions, we include SWAP gates to reverse the order of qubits as the final step in the $(\titleabbrev_{\gcds{}, \adds{}})^{\dagger}$ gate, while keeping the definition of $b_i$ unchanged, in order to more closely match the traditional definition of the $QFT$ gate.
However, to match the circuit given in the paper, i.e., without final SWAP gates, $\Tilde{u}$ (defined below) should be exchanged with $u$ in the following equations.

The $\titleabbrev_{\gcds{}, \adds{}}$ gate, given some GCDs $\gcds{}$ and additions $\adds{}$, maps a quantum state $\ket{x} = \sum_{k=0}^{2^\numqub - 1} x_k \ket{k}$ to a state $\sum_{k=0}^{2^{\numqub}-1} y_k \ket{k}$ according to the formula:
\begin{align}
    y_k = \frac{1}{\sqrt{2^{\numqub}}} \sum_{j=0}^{2^{\numqub}-1} x_j \exp[i (b \cdot j)(\Tilde{u} \cdot k)].
\end{align} 
where the $\cdot$ operation of two operands $p$ and $q$ is $p \cdot q = \sum_{i=0}^{\numqub - 1} p_i q_i$, $\Tilde{p}_i = p_{n - 1 - i}$, $q_i$ is the $i$-th bit of the binary representation of $q$, $b_{i}$ is the $i$-th bit value of the \titleabbrev\ procedure,
and $u_i$ is the number of unitary applications on the $i$-th qubit, given by
\begin{align}
    u_i = \frac{d}{\prod_{k = 0}^i G_k}.
\end{align}

When $\ket{x}$ is a basis state, the $\titleabbrev_{\gcds{}, \adds{}}$ gate can be expressed as the map
\begin{align}
    \titleabbrev_{\gcds{}, \adds{}} : \ket{x} \rightarrow \frac{1}{\sqrt{2^{\numqub}}} \sum_{k=0}^{2^{\numqub}-1} \exp[i (b \cdot x)(\Tilde{u} \cdot k)] \ket{k}.
\end{align}
The unitary matrix of the $\titleabbrev_{\gcds{}, \adds{}}$ gate acting on quantum state vectors is then
\begin{align}
    \frac{1}{\sqrt{2^\numqub}} \begin{bmatrix}
        1 & 1 & 1 & 1 & \dots & 1 \\
        1 & e^{i b_0 \Tilde{u}_0} & e^{i b_1 u_0} & e^{i (b_0 + b_1) \Tilde{u}_0} & \dots & e^{i c \Tilde{u}_0} \\
        1 & e^{i b_0 \Tilde{u}_1} & e^{i b_1 \Tilde{u}_1} & e^{i (b_0 + b_1) \Tilde{u}_1} & \dots & e^{i c \Tilde{u}_1} \\
        1 & e^{i b_0 (\Tilde{u}_0 + \Tilde{u}_1)} & e^{i b_1 (\Tilde{u}_0 + \Tilde{u}_1)} & e^{i (b_0 + b_1) (\Tilde{u}_0 + \Tilde{u}_1)} & \dots & e^{i c (\Tilde{u}_0 + \Tilde{u}_1)} \\
        \vdots & \vdots & \vdots & \vdots & & \vdots \\
        1 & e^{i b_0 \numuni} & e^{i b_1 \numuni} & e^{i (b_0 + b_1) \numuni} & \dots & e^{i c \numuni}
    \end{bmatrix},
\end{align}
where $c = \sum_{i=0}^{\numqub-1} b_i$ and $r = \sum_{i=0}^{\numqub-1} \Tilde{u}_i$.

\end{document}